\documentclass[prb,twocolumn,superscriptaddress,%
preprintnumbers,amsmath,amssymb]{revtex4}
\usepackage{amsmath}
\usepackage{graphicx}
\usepackage{dcolumn}
\usepackage{multirow}

\begin{document}

\title{Theoretical overview of black phosphorus}

\author{A. Chaves} \email{andrey@fisica.ufc.br}
\affiliation{Departamento de Fisica, Universidade Federal do Ceara, CP 6030, CEP 60455-900, Fortaleza, Ceara, Brazil} \affiliation{Department of Chemistry, Columbia University, 10027 New York, NY}  
\author{W. Ji}
\affiliation{Department of Physics, Renmin University of China, Beijing 100872, China}
\affiliation{Beijing Key Laboratory of Optoelectronic Functional Materials \& Micro-nano Devices, Renmin University of China, Beijing 100872, China.}
\author{Jesse Maassen}
\affiliation{Department of Physics and Atmospheric Science, Dalhousie University, Halifax, Nova Scotia, Canada, B3H 4R2}
\author{Traian Dumitrica}
\affiliation{Department of Mechanical Engineering, University of Minnesota, Minneapolis, MN 55455, USA}
\author{Tony Low} \email{tlow@umn.edu}
\affiliation{Department of Electrical \& Computer Engineering, University of Minnesota, Minneapolis, MN 55455, USA}

\begin{abstract}
We review the basic optical, electronic, optoelectronic, thermoelectric and mechanical properties of few-layer black phosphorus (BP), a layered semiconductor that can be exfoliated from bulk BP, the most stable allotrope of phosphorus. The distinguishing trait of BP is its highly anisotropic crystal structure, which leads to strong optical linear dichroism, elliptical plasmonics energy surfaces, anisotropic electronic and thermal conductivities and elasticity. We provide tutorial-like discussion of these phenomena and their theoretical models. Published in Avouris P, Heinz TF, Low T. \textit{2D Materials}. Cambridge University Press (2017).
\end{abstract}

\maketitle

\section{Introduction} 

	Black phosphorus (BP) is an elemental layered material composed of layers of atoms organized in a stable linked-rings structure, where the layers are held together by weak van der Waals forces, which facilitates its exfoliation into few-layers form from its bulk parent. Such atomic structure resembles the one of graphite, but the atomic rings in black-phosphorus layers exhibit a puckered structure, yielding strongly anisotropic in-plane properties, such as angle-dependent electrical and thermal conductivities, as well as excitonic polarization. Moreover, the observed high carrier mobilities of $\approx$1000 cm$^2$/Vs along its light effective mass direction makes it promising for electronics applications. \cite{5,7,1,4} 

In fact, the unique crystal structure and anisotropic properties of bulk BP had already drawn attention to this material for decades, from the 50's to the 80's. \cite{3} The interest on this material has been renewed since 2014, when its exfoliation into few-layers was recently demonstrated, \cite{5,7,1,4,6} allowing for its fabrication as an atomically thin semiconductor, whose optical band gap and electronic properties are controllable by the number of layers. For instance, the optical bandgap of BP is found to range from 0.3 eV, in bulk BP, to $\approx$ 1.6 eV, in monolayer BP,\cite{3,6,2,Liu,Rudenko1,15} making it a promising material for optoelectronics across a wide spectrum. \cite{12, Buscema, LowDevice, EngelPhotoDetector, Yuan}
	
In what follows, we discuss the theory of the electronic band structure, optical, electronic, thermal and mechanical properties of monolayer to few-layer black-phosphorus.

\section{Crystal and electronic band structures}

\begin{figure}[!h]
\includegraphics[width = \linewidth]{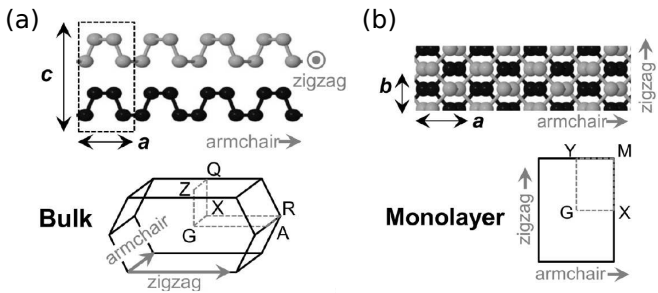}
\caption{Crystal structure of (a) bulk and (b) monolayer black phosphorus (BP) with respect to the in-plane zigzag and armchair directions. Monolayer BP corresponds to only the light or dark atoms. High symmetry points in the Brillouin zone for bulk (left) and monolayer (right) are shown in the bottom.}\label{fig.Crystal}
\end{figure}

The crystal structures of bulk and monolayer BP are shown in Figures \ref{fig.Crystal}(a) and \ref{fig.Crystal}(b), respectively, along with their first Brillouin zone schemes (bottom), where high symmetry points are highlighted. Their two main in-plane directions, namely, zigzag and armchair, are labelled. Bulk and monolayer BP have a base-centered orthorhombic [\textit{Cmce} space group (No. 64)] and simple orthorhombic [\textit{Pmna} space group (No. 53)] crystal structures, \cite{CrystTable} respectively, both with four P atoms per primitive cell. Their crystal structures are defined by the lattice vectors $\vec {a}_1 = [a \quad 0 \quad 0]$, $\vec{a}_2 = [0 \quad b/2 \quad c/2]$, $\vec{a}_3 = [0 \quad -b/2 \quad c/2]$, for bulk BP, and $\vec{a}_1 = [a \quad 0]$ and $\vec{a}_2 = [0 \quad b]$, for monolayer BP, where the distances $a$, $b$ and $c$ are represented in Figure \ref{fig.Crystal}. Examples of values for these distances, as obtained by DFT calculations \cite{15}, are $a$ = 4.47 \AA\,, $b$ = 3.34 \AA\,, $c$ = 10.71 \AA\,, for bulk BP, and $a$ = 4.58 \AA\,, $b$ = 3.32 \AA\, for monolayer BP. 

\begin{figure}[!h]
\includegraphics[width = \linewidth]{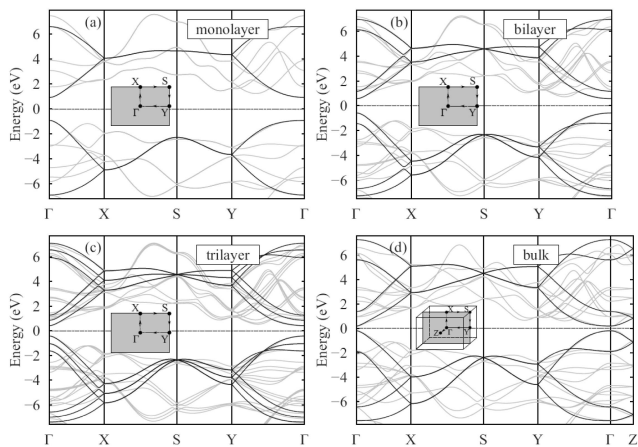}
\caption{Band structure and first Brillouin zone (insets) of few-layer and bulk black phosphorus, as obtained from GW (lighter lines) and TB (darker lines) calculations. Notice the high symmetry points are labelled differently from those in Fig. \ref{fig.Crystal}: G$\rightarrow \Gamma$, M$\rightarrow$S. After Ref. \onlinecite{10}.}
\label{fig:KatsnelsonBands}       
\end{figure}

	The electronic band structure of bulk and few-layer BP has been investigated by several groups, using \textit{ab initio} techniques with different levels of approximation, mostly using the Becke three parameters Lee-Yang-Par functional (B3LYP), \cite{6} the Perdew-Burke-Ernzerhof (PBE), \cite{15} and Heyd-Scuseria-Ernzerhof (HSE06) functionals.\cite{7,15} In general, due to the weak interlayer coupling, an appropriate treatment of van der Waals interactions in the DFT calculations is required, especially in the bulk and multi-layer cases. Particularly, monolayer BP has a predicted direct bandgap of $\sim$1.5 eV with hybridized \cite{15} or meta-GGA functionals, \cite{15, Sengupta} 0.7 eV to 1.0 eV with GGA functionals, \cite{6, Fei, 11} 1.2 to 1.6 eV (optical gap) and 2.0 to 2.3 eV (fundamental gap) for GW methods, \cite{Rudenko1, 16, 17},  and 2.15 eV with the B3LYP functional. \cite{6} Among all these various methods, GW is regarded as the most reliable one in obtaining electronic band gaps closer to experimental values.

	Figure \ref{fig:KatsnelsonBands} presents a general overview of the band structures of few-layer BP with $n$ = 1, 2, and 3 layers, as well as of bulk BP, as obtained by GW calculations.\cite{10} Dark lines are results obtained by tight-binding model, which agree very well with GW results in the vicinity of the $\Gamma$ point. The band structure for any number of layers exhibits a direct gap around the $\Gamma$ point of the first Brillouin zone. This gap decreases with the number of layers $n$ as $E_{gap} = Ae^{-Bn}/n^C + D$, with $A$ = 1.71 eV, $B$ = 0.17, $C$ = 0.73 and $D$ = 0.4 eV. \cite{10} Notice that using this expression, the bulk limit $n \rightarrow \infty$ leads to $E_{gap} \rightarrow$ 0.4 eV, very close to the well known 0.3 eV gap for bulk black phosphorus.\cite{3} We will discuss in more detail the effect of interlayer coupling and stacking on the electronic gap below.

\subsection{Monolayer black phosphorus}
\subsubsection{k . p approximation}

Let us first develop an effective Hamiltonian describing low energy bands around the G point  of the first Brillouin zone in monolayer BP in Figure \ref{fig.Crystal}(b) [or, equivalently, the $\Gamma$ point in the inset of Figure \ref{fig:KatsnelsonBands}(a)]. We shall use the $\vec k \cdot \vec p$ approximation, first presented in \cite{11}, which consists in assuming a Bloch wave function $\Psi(\vec r) = exp \left(i \vec k \cdot \vec R \right) u_k( \vec r )$ as solution for the Schr\"odinger equation describing an electron in the presence of all atoms of the BP (periodic) atomic lattice. This leads to a perturbative ($\vec k \cdot \vec p$) term $H_{\vec k \cdot \vec p} = \hbar(k_x p_x + k_y p_y)/2m_0$ and a kinetic energy $H_0 = \hbar^2 k^2/2m_0$ in the Hamiltonian, where $m_0$ is the free electron mass. We then use the atomic orbital wave functions of the unit cell as a basis to write an effective $\vec k \cdot \vec p$ Hamiltonian matrix: to lowest order, its diagonal terms contain only the conduction ($E_c$) and valence ($E_v$) quasi-particle energies. The perturbative terms involving linear dependence on the momentum can only enter the off-diagonal, since quasi-particle wave functions $| \Psi_i^s \rangle$ ($i = c$, for conduction, or $v$, for valence) are either even ($s$ = +1) or odd ($s$ = -1) with respect to $\sigma_h$ reflection, i.e. $\langle \Psi^{s}_{i} | {\hat p}_{x/y} | \Psi^{s}_{i} \rangle = 0$. Moreover, these linear terms can be inferred from symmetry arguments: since the reflection operator is unitary, $\langle \Psi^{s}_{i} | {\hat p}_{x/y} | \Psi^{s'}_{ j } \rangle = \langle \Psi^{s}_{i} | \sigma_h^{\dagger} \sigma_h {\hat p}_{x/y} \sigma_h^{\dagger} \sigma_h | \Psi^{s'}_{ j } \rangle =  \pm ss'\langle \Psi^{s}_{i} | {\hat p}_{x/y} | \Psi^{s'}_{ j } \rangle$, therefore, the matrix element for $\hat{p}_x$ ($\hat{p}_y$) is non-zero only if the states have the same (different) reflection symmetry. As from first principles calculations, the conduction and valence band states are both even with respect to $\sigma_h$ and, therefore, only $k_x$ terms appear in the off-diagonal of the effective Hamiltonian matrix. 

The second order terms in the effective Hamiltonian come from remote band contributions, within quasi-degenerate perturbation theory, a.k.a L\"owdin partition method. This method consists in finding an unitary transformation $\bar H = e^{-S}He^{S}$ on the Hamiltonian $H = H_0 + H_{\vec k \cdot \vec p }$ such that the matrix elements ${\bar H}_{mn} = \langle \Psi^s_i|{\bar H } | \Psi^r_n \rangle$ between conduction/valence states $| \Psi^s_i \rangle$ and remote band states $|\Psi^r_m \rangle$ vanish up to a desired order of $H_{\vec k \cdot \vec p}$. By constructing $S$ for a second order correction, \cite{11} one obtains
\begin{equation}
\left({\bar H}_{ \vec k \cdot \vec p } \right)_{ij} = \frac{1}{2} \sum_{l} {(H_{\vec k \cdot \vec p})_{il}(H_{\vec k \cdot \vec p})_{lj} \left[ \frac{1}{E_i - E_l} - \frac{1}{E_j - E_l} \right]}
\end{equation}
where the summation goes over the remote bands with energy $E_l$. From the form of $H_{\vec k \cdot \vec p}$,  it is clear that the diagonal and off-diagonal matrix elements may be generalized as $\left({\bar H}_{ \vec k \cdot \vec p } \right)_{ii} = {\eta_i k_x^2 + \nu_i k_y^2}$ and $\left({\bar H}_{ \vec k \cdot \vec p } \right)_{ij} = {\gamma_x k_x^2 + \gamma_y k_y^2}$, respectively. Notice that $\gamma_{x(y)}$ has contributions only from remote states with even (odd) symmetry with respect to $\sigma_h$.
	Finally, the effective k.p Hamiltonian matrix reads
\begin{equation}\label{eq.Hamkpeff}
H^{\vec k \cdot \vec p}_{eff} = \left(\begin{tabular}{cc}
$E_c + \eta_c k_x^2 + \nu_c k_y^2$ & $-i\chi k_x + \gamma_x k_x^2 + \gamma_y k_y^2$ \\
$i\chi k_x + \gamma_x k_x^2 + \gamma_y k_y^2 $ & $ E_v + \eta_v k_x^2 + \nu_v k_y^2$ \\
 \end{tabular} \right) 
\end{equation}
where the first order off-diagonal term is given by $\chi = -i \hbar \langle \Psi_c | {\hat p}_x | \Psi_v \rangle/2m_0$. The parameters  $\eta_i$, $\nu_i$, $\gamma_x$, $\gamma_y$, and $\chi$ are then chosen so to fit the electronic bands obtained by DFT. Examples of values for these parameters are provided, e.g., in Ref. \onlinecite{12}. Notice that, although the Hamiltonian Eq. (\ref{eq.Hamkpeff}) presents linear momentum terms in the off-diagonals, they are complex, so that the final energy spectrum presents no linear momentum terms, thus preserving time-reversal symmetry. \cite{13}

	In the following section, we will demonstrate that a similar low-energy Hamiltonian can also be obtained from a tight-binding model.

\subsubsection{Tight-binding model}

As an alternative to the $\vec k \cdot \vec p$ method, one can derive the approximate low energy (continuum) model Hamiltonian from the (discrete) tight-binding model Hamiltonian 
\begin{equation}\label{eq.HTB}
H_{TB} = \sum_i \varepsilon_i c_i^{\dagger}c_i + \sum_{i \neq j} \tau_{ij}c_i^{\dagger}c_j
\end{equation}
where $c_i^{\dagger}$ ($c_i$) creates (annihilates) an electron in the $i$-th atomic site, with on-site energy $\varepsilon_i$. The hopping energies $\tau_{ij}$ between the $i$-th and $j$-th sites can be obtained by comparing the tight-binding Hamiltonian $H_{TB}$ with e.g. the matrix elements of the GW Hamiltonian for the same system, written in a basis of localized Wannier functions, as first carried out in Ref. \onlinecite{Rudenko1}
	
\begin{figure}[!t]
\includegraphics[width = \linewidth]{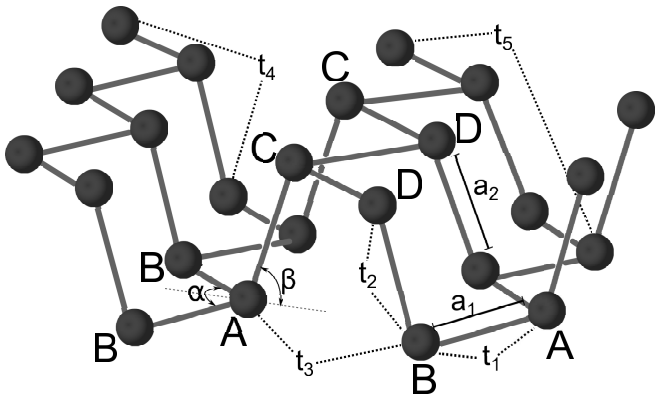}
\caption{Sketch of the atomic sites of black phosphorus. Inter-atomic distances between A and B (or C and D) atoms are $a_1$ = 2.21 \AA\,, whereas the distance between B and D (or A and C) atoms is $a_2$ = 2.24 \AA\,. Hopping energies are $t_1$ = -1.22 eV, $t_2$ = 3.655 eV, $t_3$ = -0.205 eV, $t_4$ = -0.105 eV and $t_5$ = -0.055 eV. Angles are $\alpha$ = 96.5$^{\circ}$ and $\beta$ = 71.96$^{\circ}$. \cite{14}}
\label{fig:SketchMonolayer}       
\end{figure}

The atomic sites in black phosphorus are organized as layers of staggered honeycomb lattices, as illustrated in Fig. \ref{fig:SketchMonolayer}, where the unit cell is defined by four atoms, labeled as A, B, C and D. 
The 4$\times$4 Hamiltonian matrix representing such 4-atoms unit cell is written in the basis $[\phi_A \quad \phi_B \quad \phi_D \quad \phi_C]^T$ as
\begin{equation}\label{eq.Ham4by4}
H_{TB} = \left(\begin{tabular}{cccc}
$u_A$ & $t_{AB}(\vec{k})$ & $t_{AD}(\vec{k})$ & $t_{AC}(\vec{k})$ \\
$t^{*}_{AB}(\vec{k})$ & $u_B$ & $t^{*}_{AC}(\vec{k})$ & $t_{AD}(\vec{k})$\\
$t^{*}_{AD}(\vec{k})$ & $t_{AC}(\vec{k})$ & $u_C$ & $t_{AB}(\vec{k})$\\
$t^{*}_{AC}(\vec{k})$ & $t^{*}_{AD}(\vec{k})$ & $t^{*}_{AB}(\vec{k})$ & $u_D$
 \end{tabular} \right) 
\end{equation}
where the structure factors
\begin{widetext}
\begin{eqnarray}
t_{AB} = 2 \cos\left(k_x a_1 \sin\frac{\alpha_1}{2}\right)\left[t_1e^{-ik_y a_1 \cos{\frac{\alpha_1}{2}}} + t_3 e^{ik_y \left(a_1 \cos{\frac{\alpha_1}{2}} + 2a_2 \cos\beta\right)}\right]   \nonumber\\
 t_{AC} = t_2 e^{i k_y a_2 \cos\beta} + t_5e^{-ik_y \left(2a_1 \cos{\frac{\alpha_1}{2}} + a_2 \cos\beta\right)}  \nonumber\\
 t_{AD} = 4t_4 \cos\left(k_x a_1 \sin\frac{\alpha_1}{2}\right)\cos\left[{k_y \left(a_1 \cos{\frac{\alpha_1}{2}} + a_2 \cos\beta\right)}\right] 
\label{eq:StructureFactors}
\end{eqnarray}
\end{widetext}
are obtained by a Fourier transform of Eq. (\ref{eq.HTB}) using the lattice vectors. Due to the symmetry of the lattice, the on-site energies are all equivalent and set to zero ($u_{A,B,C,D} = 0$) from here onwards.

The Hamiltonian Eq. (\ref{eq.Ham4by4}) can be further simplified by symmetry arguments: one can couple wave functions for A and D sites, and for B and C sites, leading to a more simple two-component spinor basis $1/2[ \phi_A+\phi_D \quad \phi_B + \phi_C]^T$. In this basis, the Hamiltonian becomes
\begin{equation}
H_{TB} = \left(\begin{tabular}{cc}
$t_{AD}(\vec{k})$ & $t_{AB}(\vec{k})+ t_{AC}(\vec{k})$ \\
$\left[t_{AB}(\vec{k})+t_{AC}(\vec{k})\right]^{*}$ & $t_{AD}(\vec{k})$\\
 \end{tabular} \right) 
\end{equation}
By expanding the structure factors around the lowest energy ($k = 0$) $\Gamma$-point and keeping only terms up to second order in $k$, one can finally write an effective low-energy Hamiltonian
\begin{equation} \label{eq.HTBeff}
H^{TB}_{eff} = \left(\begin{tabular}{cc}
$\delta_0 + \eta k_x^2 + \nu k_y^2$ & $\delta_1 + i\chi k_y + \gamma_x k_x^2 + \gamma_y k_y^2$ \\
$\delta_1 -i\chi k_y + \gamma_x k_x^2 + \gamma_y k_y^2$ & $ \delta_0 + \eta k_x^2 + \nu k_y^2$\\
 \end{tabular} \right) 
\end{equation}
where
\begin{eqnarray}\label{eq.TBparameters}
\delta_0 = 4t_4 \nonumber\\ 
\eta = - 2t_4[a_1\sin(\alpha_1/2)]^2 \nonumber\\
\nu =-2t_4[a_1\cos(\alpha_1/2) + a_2\cos\beta]^2 \nonumber\\
\gamma_x =-(t_1+t_3)[a_1 \sin(\alpha_1/2)]^2 \nonumber\\
\gamma_y= - t_1[a_1\cos(\alpha_1/2)]^2 - t_3[a_1\cos(\alpha_1/2) + 2a_2\cos\beta]^2 - \nonumber\\
- t_2(a_2\cos\beta)^2/2-t_5[2a_1\cos(\alpha_1/2) +a_2\cos\beta]^2/2 \nonumber\\
\delta_1=t_2+t_5+2(t_1+t_3) \nonumber\\
\chi = 2(t_3-t_5-t_1)a_1 \cos \left(\alpha_1/2 \right) + (4t_3 - t_5 + t_2) a_2 \cos\beta.
\end{eqnarray}
Notice the high similarity between the tight-binding and the $\vec k \cdot \vec p$ Hamiltonians, Eq. (\ref{eq.HTBeff}) and Eq. (\ref{eq.Hamkpeff}), respectively. Their basic differences are on the constants, the off-diagonal linear terms (which are in the y-direction in $H^{TB}_{eff}$ ) and on the fact that the diagonal second-order factors $\eta$ and $\nu$ are not band dependent in the tight-binding Hamiltonian. Nevertheless, the tight-binding model brings the advantage of providing an estimate of the value of all these parameters, without the need to fit the band structures previously obtained by DFT. By substituting the values of angles and hoping parameters in Eq. (\ref{eq.TBparameters}), one obtains $\eta = 0.58$ eV\AA\,$^2$, $\nu = 1.01$ eV\AA\,$^2$, $\gamma_x = 3.93$ eV\AA\,$^2$, $\gamma_y = 3.83$ eV\AA\,$^2$, $\delta_0 = -0.42$ eV, $\delta_1 = 0.76$ eV, and $\chi = 5.25$ eV\AA\,.

\subsubsection{Effective mass}

Diagonalization of the 2$\times$2 Hamiltonian matrix Eq. (\ref{eq.HTBeff}) [or, equivalently, Eq. (\ref{eq.Hamkpeff})] leads to 
\begin{equation}
E(\vec{k}) = \delta_0 + \eta k_x^2 + \nu k_y^2 \pm \sqrt{(\delta_1 +\gamma_x k_x^2+ \gamma_y k_y^2)^2 + \chi^2 k_y^2}                                   ,                                      
\end{equation} 
where the + (-) sign refer to the conduction (valence) band. These bands are clearly non-parabolic and anisotropic. However, by expanding the square-root term in this equation up to second order in $\vec k$ around $k = |\vec k| = 0$, one obtains approximate parabolic bands
\begin{equation}
E_{\pm} = ( \delta_0 + \delta_1 ) + \frac{\hbar^2 k_x^2}{ 2m^{\pm}_x }+ \frac{\hbar^2 k_y^2}{2m^{\pm}_y },		
\end{equation}			              
This allows us to obtain expressions for effective masses for conduction and valence band states in BP as
\begin{eqnarray}\label{eq.mass}
m_x^{+} = \frac{\hbar^2}{2(\gamma_x + \eta)} = 0.846 m_0 \\
m_x^{-} = \frac{\hbar^2}{2(\gamma_x - \eta)} = 1.14 m_0 \\
m_y^{+} = \frac{\hbar^2}{2(\gamma_y + \nu + \chi^2/2\delta_1)} = 0.166 m_0 \\
m_y^{-} = \frac{\hbar^2}{2(\gamma_y - \nu - \chi^2/2\delta_1)} = 0.182 m_0 
\end{eqnarray}

\begin{figure}[!h]
\includegraphics[width = \linewidth]{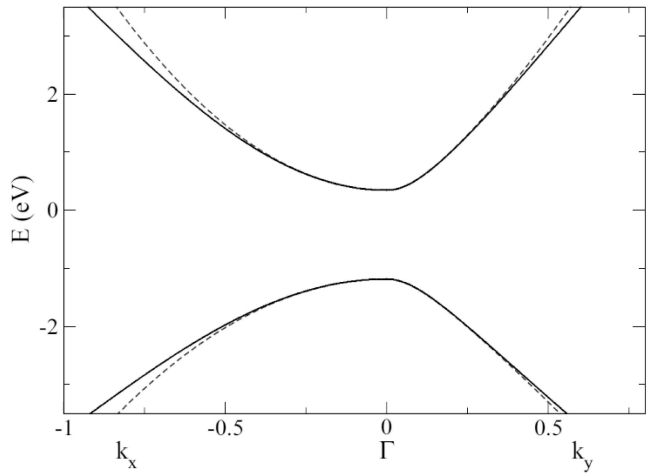}
\caption{Comparison between energy bands around the $\Gamma$ point obtained by the full tight-binding Hamiltonian (solid lines) and parabolic (dashed lines) models. After Ref. \onlinecite{14}.}
\label{fig:ComparisonEffMass}       
\end{figure}

The energy bands obtained by the full tight-binding Hamiltonian and the parabolic approximation are shown in Fig. \ref{fig:ComparisonEffMass}, for comparison.

\subsection{Interlayer coupling and stacking}

\begin{figure}[!h]
\includegraphics[width = \linewidth]{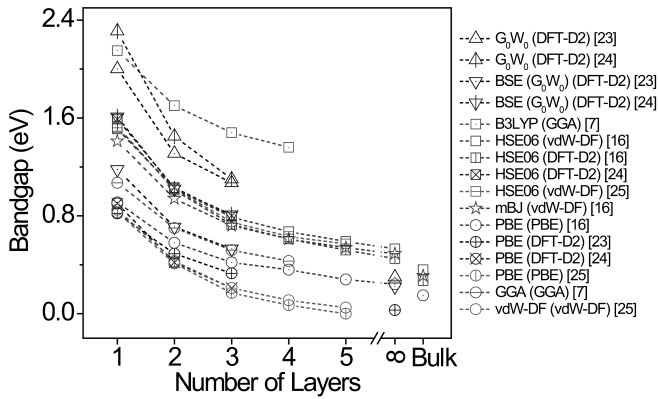}
\caption{Electronic bandgap of few-layer BP as a function of layer thickness, as predicted by DFT calculations with different functionals. \cite{6,15,16,17,18}}
\label{fig:Gaps}       
\end{figure}

The bandgap of few-layer BP is of particular interest, since it is drastically reduced by adding only one additional layer. In fact, few-layer BP has the strongest layer-dependent modulation of electronic bandgap among all 2D materials being reported so far. Figure \ref{fig:Gaps} summarizes the predicted bandgap as a function of the thickness of BP layers, which, as previously discussed, shows an exponential decay as the thickness increases, regardless which functional was used in the calculations. Interestingly enough, the bandgap numerically extrapolated to the case of infinite number of layers is slightly larger than that calculated from a real bulk sample, which is a result of the shortened lattice constant in the few-layers case.

\begin{figure}[!h]
\includegraphics[width =0.6 \linewidth]{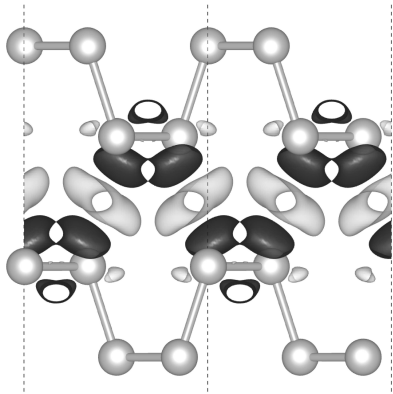}
\caption{Differential charge density at the interlayer region of 6L-BP. The electronic charge density is derived by computing $\rho_{6L-BP} - \rho_{3L-BP_u} - \rho_{3L-BP_d}$, where $\rho_{6L-BP}$ is the total charge density and $\rho_{3L-BP_u}$ and $\rho_{3L-BP_d}$ represent the total charge densities of the upper and lower halves of the system, respectively. Light-colour contours indicate charge accumulation, whereas dark-colour ones stand for charge reduction. \cite{19}}
\label{fig:6Layers}       
\end{figure}

Interlayer coupling plays a critical role in the resulting bandgap evolution of BP. Its differential charge density [Figure \ref{fig:6Layers}] explicitly shows a covalent-like characteristic in the inter-layer region of few-layer BP, in which charge reduction was found near the P atoms and charge accumulation was observed in-between the inter-layer P atoms. If we adopt a molecular orbital language for a simple analysis, each P atom takes the $sp^3$ hybridization, thus forming three covalent bonds and leaving along one electron pair pointing almost along the out-of-plane direction. The interlayer van der Waals attraction is so strong that it brings the two BP layers close enough for having the two electron pairs of the two P atoms from the approaching layers being forced to hybridize, thus showing a bonding and an anti-bonding states. Such substantial electronic hybridization distinguishes the interlayer coupling of BP from that of other 2D materials discussed in this book. In order to stand out this specialty, the term ``covalent-like BP quasi-bonding" has been proposed to denote this vdW attraction induced electronic hybridization. \citep{19}

\begin{figure}[!h]
\includegraphics[width = \linewidth]{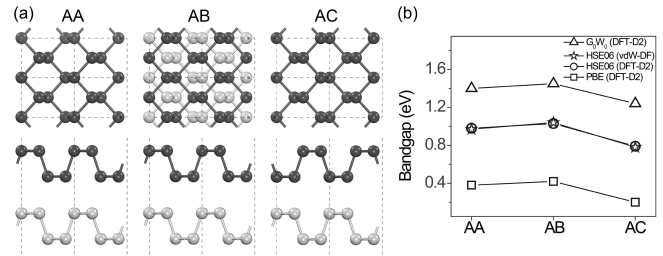}
\caption{Atomic configuration and predicted bandgap of three stacking orders for bilayer BP. (a) Top and side views of the AA-, AB- and AC-stacked bilayer BP. (b) Bandgaps of different stacking orders calculated with the PBE \cite{17} and HSE06 functionals \cite{17, 20}, as well as with GW approximation \cite{17}.}
\label{fig:ManyLayers}       
\end{figure}

Electronic bandgaps of few-layer BP are largely determined by the interlayer electronic hybridization, which can be affected by the interlayer distance and the stacking order between two layers. There are, at least, three types of different stacking orders, namely AA, AB and AC, see Fig. \ref{fig:ManyLayers}(a). The AB-stacked BP is the most widely investigated and has the lowest energy among the three. \cite{17,20} Results previously presented in Fig. \ref{fig:KatsnelsonBands} were obtained assuming this type of stacking. If we laterally shift one layer of a bilayer BP, the bandgap varies by 0.2 - 0.4 eV, as predicted by different methods and shown in Fig. \ref{fig:ManyLayers}(b), due to the change of interlayer hybridization. The variation of optical bandgaps become even larger if more layers are stacked on a bilayer BP. For instance, a $\sim$0.6 eV change of optical bandgap has been reported for an ACA stacked tri-layer BP with the $G_0W_0$ method. Nevertheless, the exciton binding energy is almost unchanged among different stacked configurations. \cite{17} 

\section{Electronic properties}

In this section, we discuss some basic electronic properties of BP, highlighting its highly anisotropic aspects, which make it uniquely different from conventional 2D electron gas systems.

\subsection{In-plane polarizability and screening}

It is well known that the polarizability of a conventional 2D electron gas is given by its density of states. The dielectric function of an electron gas in the random phase approximation (RPA) can then be written in terms of its polarizability $\Pi$ as \cite{21},
\begin{equation}
\epsilon \left(\vec{q},\omega \right) =\kappa + {v}_{c} \left( q \right)\Pi(\vec{q},\omega),
\end{equation}
where ${v}_{c}\left(q \right) = {{e}^{2}} \big/ {2 {\epsilon}_{0}q}$ is the 2D Coulomb interaction, $\vec q$ is the plasmon wave-vector, $\epsilon_0$ is the permittivity of free space and $\kappa$ the effective dielectric constant of its environment. $\Pi(\vec q, \omega)$ is the 2D dynamic polarizability, whose expression can be found in standard textbooks on 2D electron gas system \citep{22}. In an anisotropic medium such as BP, $\Pi(\vec q, \omega)$ in general is also anisotropic and depends on the direction of $\vec q$, a quantity that has to be computed numerically \cite{23}. 

In the static limit, however, it is possible to obtain a closed form expression for the BP polarizability. We generalize the well-known analytical form of the polarizability for 2D electron gas to include anisotropy \cite{21}. Since $\omega = 0$, we deal only with intraband processes. In the $T = 0$ limit, we obtain \cite{23},
\begin{equation}
\Pi \left(\vec {q} \right) =  {g}_{2D} \Re \left[1-\sqrt{1-\frac{8\mu / {\hbar}^{2}} {{{q}_{x}^{2}} / {m}_{x} + {{q}_{y}^{2}} /{m}_{y}}} \right]
\end{equation}
where ${g}_{2D} = {{m}_{d}}/{\pi {\hbar}^{2}}$ is the 2D density-of-states, $m_d = \sqrt{m_x m_y}$, and $\mu$ is the chemical potential. We make an interesting observation: for $q \leq | {\vec {k}}_{F} \cdot \vec {q}|$, we see that $\Pi(\vec{q})$ reduces to the familiar relation for the static polarizability of an isotropic 2D electron gas, $\Pi(\vec{q}) = g_{2D}$. \cite{21} In other words, for long-range potentials, such as those induced by charged impurities, which involve momenta in accordance with $q \leq | {\vec {k}}_{F} \cdot \vec {q}|$, the screening will be isotropic, at least in the zero temperature and disorder limits.

\subsection{Plasmons}

Plasmons in 2D materials has attracted significant attention recently as they provide an interesting platform for enhanced light-matter interactions \cite{24,25, Bludov, Koppens}. In an isotropic media like graphene, the in-plane plasmon dispersion just follows the simple relation ${\omega}_{pl} \left(q \right) = \sqrt{{Dq} \big/ {2\pi {\epsilon}_{0} \kappa}}$, where $\vec q$  is the plasmon wave-vector, $D = \pi e^2 n_e/m$ is the Drude weight, $m$ is the diagonal mass tensor for BP, and $n_e$  is the electron density \cite{24, Mikhailov}.

Here, we discuss the plasmon dispersion in an anisotropic media like BP \cite{23}. Consider a BP film in the $xy$-plane sandwiched between two dielectric media $\epsilon_1$ and $\epsilon_2$. We are interested in the bound electromagnetic modes, characterized by an in-plane wave-vector $\vec{q}$ pointing at an angle $\theta$ with respect to the $x$-axis. From Maxwell equations, one can show that the bound mode has to satisfy the following equation,
\begin{equation}\label{eq.plasmonsfull}
 \left({\acute{Y}}_{s} + {M}_{ss}\right) \left ({\acute {Y}}_{p} + {M}_{pp} \right) - {M}_{ps} {M}_{sp} = 0,
\end{equation}
where ${\acute {Y}}_{\beta} = {Y}_{\beta}^{1} + {Y}_{\beta}^{2}$ ($\beta = s, p$) is the total admittance, with ${Y}_{s}^{i} = {Y}_{0} ( {{k}_{zi}} / {{k}_{0}} )$  and ${Y}_{p}^{i} = {Y}_{0} {\epsilon}_{i} ( {{k}_{0}} / {{k}_{zi}} )$, and ${k}_{zi}^{2} = {k}_{0}^{2} {\epsilon}_{i} - {q}^{2}$, $k_0 = \omega/c$. $c$ and $Y_0 = \sqrt{\epsilon_0/\mu_0}$ are the speed of light and admittance of free space, respectively. The matrix elements for $M$ are given by,
\begin{eqnarray}
{M}_{ss} = {\sigma}_{xx} \sin^2 \theta+ \sigma_{yy}\cos^2\theta \nonumber\\
{M}_{pp} = {\sigma}_{xx}\cos^2\theta+ {\sigma}_{yy}\sin^2\theta \nonumber \\
{M}_{sp} = {{M}_{ps}} = ({\sigma}_{xx} - {\sigma}_{yy})\sin\theta\cos\theta
\end{eqnarray}                                                       
where $\sigma_{ij}$ is the 2D conductivity tensor. In the simple case where $\theta = 0,\pi$ and $\sigma_{xx} = \sigma_{yy}$, Eq. (\ref{eq.plasmonsfull}) reduces to
\begin{equation}
{\acute{Y}}_{p} + {M}_{pp} =0
\end{equation}

In the non-retarded regime, i.e. $q \gg k_0$ , hence $k_{zi} \approx iq$, we obtain the 'quasi-static' approximation,
\begin{equation}\label{eq.plasmoncondition}
- \frac{{\sigma}_{xx}\cos^2\theta + {\sigma}_{yy}\sin^2\theta}{{\epsilon}_{0}\omega} = \frac{{\epsilon}_{1}}{{k}_{z1}} + \frac{{\epsilon}_{2}}{{k}_{z2}} \approx \frac{2\kappa}{iq},
\end{equation}	
where $\kappa = (\epsilon_1 + \epsilon_2)/2$. An often used expression for the conductivity is given by the Drude model,
\begin{equation}\label{eq.ConductivityDrude}
{\sigma}_{jj} (\omega)= \frac{i {D}_{j}}{\pi(\omega+i\delta)}
\end{equation}
where $D_j = \pi e^2 n_e/m_{jj}$  is the Drude weight, $m_{jj}$ is the diagonal mass tensor for BP, $n_e$ is the electron density, and $\delta$ is the single particle damping rate. Unlike the isotropic case, the Drude weight is now directional dependent. Hence, Eq. (\ref{eq.plasmoncondition}) in conjunction with Eq. (\ref{eq.ConductivityDrude}) will provide the dispersion of the plasmon in an anisotropic media like BP.

\subsection{Electrical Transport}

As a consequence of its anisotropic electronic structures, the carrier mobility and electronic transport properties in BP are highly directional. In general, the mobility might be reduced by phonon, impurity and interface scatterings. Among them, the phonon scattering is intrinsic at room temperature, but the impurity scattering could be largely suppressed by reducing defect concentration and an optimized metal-semiconductor or semiconductor-insulator interface structure could highly improve the interface scattering. The phonon limited mobility is determined by the scattering of carriers with both elastic acoustic and inelastic optical phonon scattering processes. At room temperature, however, elastic acoustic scattering process is dominant, therefore, the discussion in the rest of this part is thus focused on the elastic acoustic phonon-carrier scattering process. 

Theory predicts rather high hole mobilities in free-standing pristine mono- to five-layer BP along the armchair ($x$) direction, while those along the zigzag ($y$) direction are times smaller, and so is the electron mobility. The hole mobility increases from a few hundreds up to several thousands cm$^2$/Vs with respect to the enlarging thickness, based on an elastic acoustic phonon-carrier scattering model.\cite{15} The difference between mobilities in different directions can be partially explained by the anisotropic carrier effective masses: both hole and electron effective masses along the $\Gamma$-$X$ direction in the $k$ space (associated to the $x$ direction in real-space) are roughly ten times smaller than those along the $\Gamma$-$Y$ ($y$) direction, namely $\approx$0.15 $m_0$ versus $\approx$1.0 $m_0$ in the monolayer case, as discussed in previous sections. At the extreme case, namely hole effective masses of monolayer, the factor of difference is over $\approx$40, namely $\approx$0.15 $m_0$ versus 6.35 $m_0$, as calculated by DFT. One may notice, though, that these values are several times larger than the experimentally measured mobility anisotropy ratio \cite{1,Mishchenko}. As a matter of fact, the impurity scattering and the interfacial Coulomb scattering are, most likely, sufficient to reduce the mobility of BP by an order of magnitude, namely down to hundreds of cm$^2$/Vs in few layers, consistent with experimental values.

\onecolumngrid

\begin{table}[!h]
\caption{Predicted carrier mobility due to electron-phonon scattering for few-layer BP. Carrier types $e$ and $h$ denote electron and hole, respectively. $N_L$ represents the number of layers, $m_x$ and $m_y$ are carrier effective masses (in units of $m_0$) for directions $x$ and $y$, respectively. \cite{17,18,26,27,28} $E_{1x}$ ($E_{1y}$) (in eV), and $C_{x2D}$ ($C_{y2D}$) (in Jm$^{-2}$), are the deformation potential and 2D elastic modulus for the $x$ ($y$) direction, respectively. Mobilities $\mu_{x2D}$ and $\mu_{y2D}$ (in 10$^3$cm$^2$V$^{-1}$s$^{-1}$) were calculated using Eq. (1.22) with the temperature $T$ set to 300 K. Adapted from Ref. \onlinecite{15}.} \label{tab:1}\renewcommand{\arraystretch}{1.5}
\vspace{1mm}
\setlength{\tabcolsep}{10pt}
\renewcommand\arraystretch{1.2}
\scriptsize
\begin{tabular}{cccccccccccccccccccccccc}
\hline
\hline
\ & ${N}_{L}$ & ${m_{x}}$ & ${m_{y}}$ & ${E_{1x}}$ & ${E_{1y}}$ & ${C_{x2D}}$ & ${C_{y2D}}$ & ${\mu_{x2D}}$& ${\mu_{y2D}}$ \\
\hline
\multirow{6}{*}{e} & 1 & 0.14 $\thicksim$ 0.23 & 1.12 $\thicksim$ 1.25& 2.72 $\pm$ 0.02 & 7.11 $\pm$ 0.02 & 28.94 & 101.60 & 1.10$\thicksim$1.14& $\thicksim$0.08\\
&2 & 0.11 $\thicksim$ 0.18 & 1.13 $\thicksim$ 1.35& 5.02 $\pm$ 0.02 & 7.35 $\pm$ 0.16 & 57.48 & 194.62 & $\thicksim$0.60& 0.14$\thicksim$0.16\\
&3 & 0.06 $\thicksim$ 0.16 & 1.15 $\thicksim$ 1.35& 5.85 $\pm$ 0.09 & 7.63 $\pm$ 0.18 & 85.86 & 287.20 & 0.76$\thicksim$0.80 & 0.20$\thicksim$0.22\\
&4 & 0.04 $\thicksim$ 0.16 & 1.16 $\thicksim$ 1.33& 5.92 $\pm$ 0.18 & 7.58 $\pm$ 0.13 & 114.66 & 379.58 & 0.96$\thicksim$1.08 & 0.26$\thicksim$0.30\\
&5 & 0.15                  & 1.18 $\thicksim$ 1.33& 5.79 $\pm$ 0.22 & 7.35 $\pm$ 0.26 & 146.58 & 479.82 & 1.36$\thicksim$1.58 & 0.36$\thicksim$0.40\\
&bulk&0.12 &1.15\\
\\
\multirow{6}{*}{h} & 1 & 0.13 $\thicksim$ 0.15 & 6.17 $\thicksim$ 13.09 & 2.50 $\pm$ 0.06 & 0.15 $\pm$ 0.03 & \  & \  & 0.64$\thicksim$0.70 & 10$\thicksim$26\\
&2 & 0.10 $\thicksim$ 0.15 & 1.81 $\thicksim$ 2.70& 2.45 $\pm$ 0.05 & 1.63 $\pm$ 0.16 & \  &\   & 2.6$\thicksim$2.8& 1.3$\thicksim$2.2\\
&3 & 0.06 $\thicksim$ 0.15 & 1.12 $\thicksim$ 1.55& 2.49 $\pm$ 0.12 & 2.24 $\pm$ 0.18 &  \ &  \ & 4.4$\thicksim$5.2 & 2.2$\thicksim$3.2\\
&4 & 0.04 $\thicksim$ 0.14 & 0.97 $\thicksim$ 1.11& 3.16 $\pm$ 0.12 & 2.79 $\pm$ 0.13 &\   &\   & 4.4$\thicksim$5.2 & 2.6$\thicksim$3.2\\
&5 & 0.14                  & 0.89 $\thicksim$ 1.01& 3.40 $\pm$ 0.25 & 2.97 $\pm$ 0.18 &\   & \  & 4.8$\thicksim$6.4 & 3.0$\thicksim$4.6\\
&bulk&0.11 &0.71\\
\hline
\hline
\end{tabular}
\end{table}

\twocolumngrid

The acoustic phonon limited carrier mobility in 2D is given by the expression                                             			      
\begin{equation}\label{eq.mobility}
\mu_{2D} = \frac{e \hbar^3C_{2D}}{k_B T m_i m_d (E_1^i)^2},
\end{equation}
where $m_i$ is the carrier's effective mass in the transport direction ($i = x, y$) and $m_d = \sqrt{m^{\pm}_x m^ {\pm}_y}$  is the effective mass averaged in different directions. The term $E_1^i = \Delta V_i/( \Delta l/l_0)$ represents the deformation potential constant of the VBM (hole) or CBM (electron) along the transport direction, where $\Delta V_i$ is the energy change of the $i$-th band under proper cell compression and expansion (calculated using a step of 0.5\%), $l_0$ is the lattice constant in the transport direction and $\Delta l$ is the deformation of $l_0$. The elastic modulus $C_{2D}$ of the longitudinal strain in the propagation directions (both $x$ and $y$) of the longitudinal acoustic wave is derived from $(E - E_0)/S_0 = C( \Delta l/l_0 )^2 / 2$, where $E$ is the total energy and $S_0$ is the lattice volume at equilibrium for a 2D system. The resulting carrier mobilities are shown in Table \ref{tab:1}.

\section{Optical properties}
\subsection{Optical absorption}

\begin{figure}[!h]
\includegraphics[width = \linewidth]{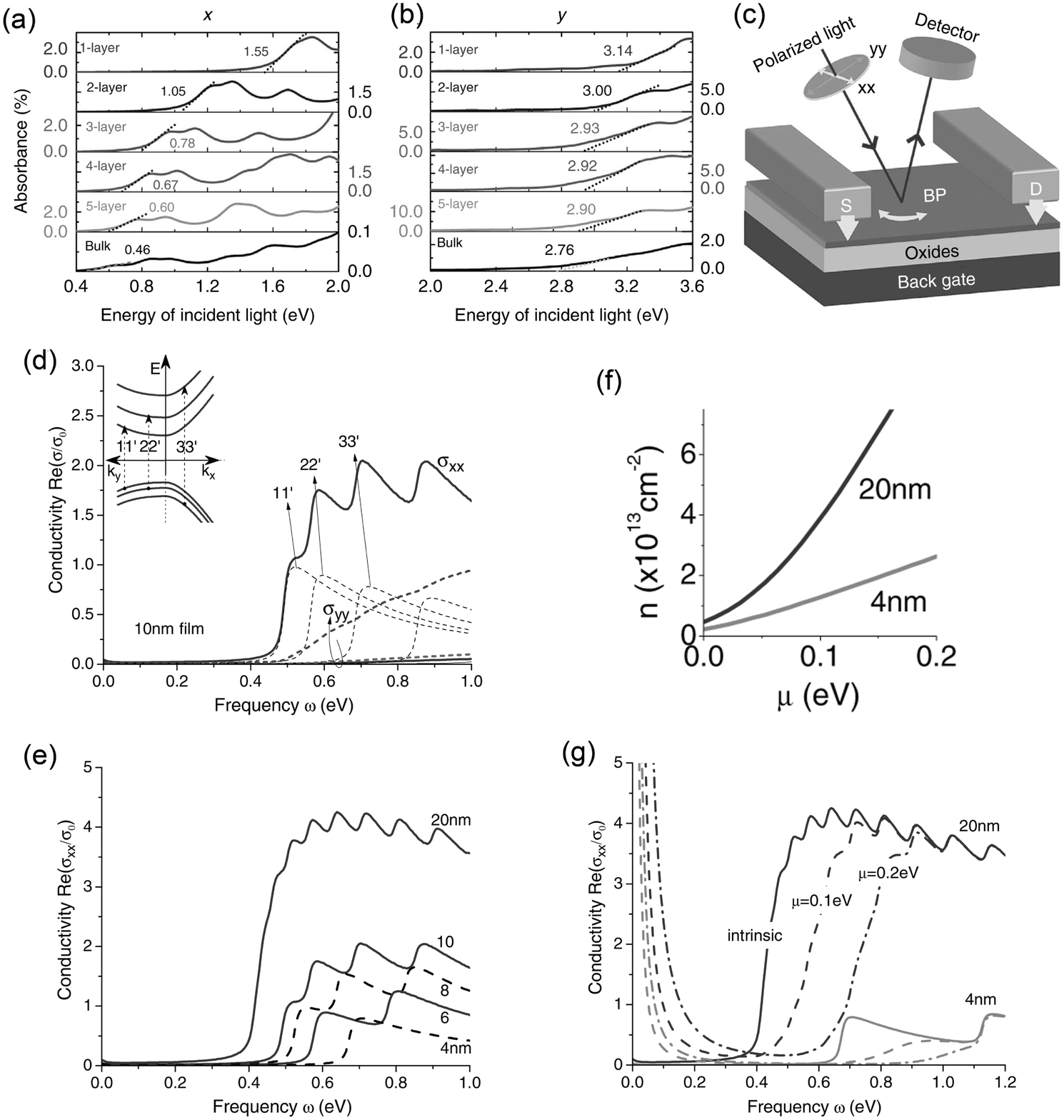}
\caption{(a-b) Optical absorption spectra of few-layer BP for light incident in the $c$ ($z$) direction and polarized along the ($a$) $x$ and ($b$) $y$ directions. (c) Schematic illustration of a proposed experimental geometry to determine the orientation of few-layer BP structures using optical absorption spectroscopy. Adapted from Ref. \onlinecite{15}. Real part of optical conductivities along the $x$ and $y$ directions [$\Re(\sigma_{xx})$ and $\Re(\sigma_{yy})$] for intrinsic BP at 10-nm-thick (d) and the former [$\Re(\sigma_xx)$] at different thicknesses (e). (f) Electron densities $n$ as function of chemical potential $\mu$ for 20- and 4-nm BP films and (g) $\Re(\sigma_{xx})$ for BP with different chemical
potential $\mu$ for these values of layer thickness.\cite{12} }
\label{fig:OpticalSpectrum}       
\end{figure}

BP exhibits a strong anisotropy for light absorption. Within a wide energy range, including infrared light and part of visible light, few-layer black phosphorous absorbs light polarized along the structures' armchair direction and is transparent to light polarized along the zigzag direction. Such light-absorption anisotropy is a result of the underlying symmetries of wave functions for the valence and conduction bands, which governs the dipole transitions and thus optical absorption. The dipole transition along the $y$ direction ($\Gamma$-Y in the $k$ space) around the bandgap, namely from VBM to CBM, is largely suppressed because of the similarity of the shape of wave functions of the valence and conduction bands along this direction. Figure \ref{fig:OpticalSpectrum} shows the absorption spectra of few-layer (a,b) and multi-layer BP (d,e) with respect to light polarization, \cite{15} thickness \cite{12,15} and charge doping, \cite{15} respectively. Taking advantage of the observed highly polarized light absorption, a possible experimental setup for determining the orientation of few-layer BP structures is proposed in Fig. \ref{fig:OpticalSpectrum}(c).\cite{12}
 
	In addition to the polarization dependent light absorption, the absorption spectra of multilayer black phosphorus are shown to vary sensitively with thickness and doping. As we discussed in previous Sections, the (direct) bandgap decreases with the thickness. Thus, as the thickness increases, the band edge reflected in the absorption spectra moves towards lower energies and eventually reaches 0.3 eV (for bulk BP). It is interesting to point out that the absorption spectra also exhibits peak like structures. These peaks are a consequence of the anisotropic interband coupling and dispersion, where each peak corresponds to conduction to valence band transitions between subbands of same indices \cite{Zhang, 12} Charge doping, modeled with a rigid shift in the chemical potential $\mu$, as shown in Fig. \ref{fig:OpticalSpectrum}(f), can be realized by chemically introducing charge donor or acceptor impurities. Extra electrons shift the band edge in the absorption spectra towards higher energies, as verified in Fig. \ref{fig:OpticalSpectrum})(g) because of the Pauli blocking of the extra charges to the direct optical transition. \cite{12} Ideally, this blue-shift is of 2$\mu$, but the peak is usually also smeared out in energy due to finite-temperature effects. Similar modulation of absorption spectra can also be achieved via electrical gating, where in this case, the Franz-Keldysh effect can also dominate. \cite{LinC, PengR} 

In the above-mentioned, optical transitions are described within the single particle picture. However, many body effects, such as excitons, usually dominates the optical spectra especially at low temperatures. \cite{Review} In what follows, we will describe how to predict the position of the lower energy peak in absorption spectrum, namely, the optical gap, as well as other excited state peaks, that would show up in low-temperature reflectance measurements.	
	
\subsection{Excitons}

In light mediated experiments in semiconductors, electrons are either excited from the valence band to the conduction band, leaving a positively charged quasi-particle (hole) in the former, as in absorption and reflectance measurements, or recombined with holes in the valence band as to produce photoluminescence. Either way, the energy involved in these processes is easily detectable as peaks in the absorption and photoluminescence spectra. The lowest energy peak is the so-called optical gap, which consists of the bare quasi-particle gap between the top of the valence band and the bottom of the conduction band, deducted by the binding energy between the electron and hole involved in the process. Remarkably high electron-hole (exciton) binding energies is a feature shared by all 2D materials, such as transition metal dichalcogenides and few layer black phosphorus. \cite{Berkelbach, 31, Tran} For WS$_2$, for example, a 320 meV exciton binding energy has been experimentally observed by reflectance experiments, which also revealed a very clear set of peaks that were interpreted as a Rydberg series of excited excitonic states, in a clear analogy between the bound electron-hole pair and a planar hydrogen atom.\cite{Alexey} As for black phosphorus, photoluminescence measurements have consistently suggested binding energies of the order of hundreds meV,\cite{2, 6} and even trions (charged exciton complexes, i.e. with an extra electron or hole) with binding energies of tens of meV. \cite{TrionBP}

\begin{figure}[!h]
\includegraphics[width = \linewidth]{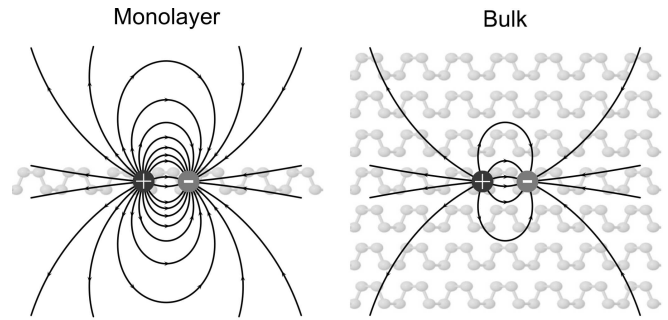}
\caption{Sketch of electron-hole pairs (excitons) in bulk and monolayer BP. The changes in the dielectric environment are schematically indicated by different spacing between the electric field lines.}
\label{fig:Exciton1}       
\end{figure}

Such high binding energies are due to the fact that, unlike in bulk semiconductors, the electron-hole Coulomb interaction in theses materials undergo a much lower screening by the vacuum surrounding the sample, or by the substrate below it. This situation is illustrated in Fig. \ref{fig:Exciton1}, where the different screening factors are represented by different separation between electric field lines. Notice that the different dielectric media can only play a role on the overall interaction potential for small electron-hole separations, when electric field lines are distributed along both vacuum and the layer. However, as the electron-hole separation increases in the quasi-two-dimensional (monolayer) case, most of the electric field lines end up outside the semiconductor, so that the contribution of the semiconductor layer is smeared out and the Coulomb interaction potential is expected to be recovered. This suggests an effective dielectric constant that depends on the electron-hole separation, which turns out to be the case, as we will demonstrate in what follows.
	
\subsubsection{Screened electron-hole interaction potential}

The actual interaction potential for charged particles in a layer surrounded by different dielectric constants media was firstly derived by Keldysh in the 70's, \cite{29} and recently revisited by several papers in the literature in the context of 2D semiconductors. \cite{30, CastroNetoExciton} In simple words, Poisson equation is solved to obtain the electrostatic potential $\Phi(\vec r)$ induced by a point charge in a infinitely thin slab of polarizability $\alpha_{2D}$, placed at $z = 0$, surrounded by vacuum:
\begin{equation}\label{eq.PoissonIni}
\nabla^2\Phi = -4\pi \left[e \delta(\vec r) - \vec \nabla \cdot {\vec P}_{2D} \right]
\end{equation}
The first term between brackets represent the point charge, whereas the second term is an induced charge density, which, in the long wavelength limit, depends on the macroscopic polarization $\vec{P}_{2D} = -\alpha_{2D}\vec{\nabla}\Phi$. Substituting in Eq. (\ref{eq.PoissonIni}) leads to
\begin{equation}
\nabla^2\Phi = -4\pi \left[e \delta(\vec r) + \alpha_{2D}\nabla^2 \Phi|_{z = 0}\delta(z) \right]
\end{equation}
which, after Fourier transform, can be re-written in reciprocal space as
\begin{equation}\label{eq.phi1}
\Phi(\vec q, k_z) =  -4\pi \left[\frac{e}{(|\vec q|^2 + k_z^2)} -\frac{\alpha_{2D}|\vec q|^2}{(|\vec q|^2 + k_z^2)}\Phi_{2D}(\vec q)\right],
\end{equation}
where we defined
\begin{equation}
\Phi_{2D} (\vec q)= \frac{1}{2\pi}\int_{0}^{\infty}dk_z \Phi(\vec q,k_z ).
\end{equation}
Notice $\Phi_{2D}(\vec q)$ is exactly the in-plane electrostatic potential we are looking for. Equation (\ref{eq.phi1}) is then integrated again in $k_z$, and terms are re-arranged as
\begin{equation}\label{eq.NonConstDielectric}
\Phi_{2 }(\vec q ) = \frac{ 2\pi e }{ |\vec q | \left(1 + 2 \pi \alpha_{2D} |\vec q| \right)}.
\end{equation}

In fact, the in-plane potential $\Phi_{2D}(\vec q)$ in Eq. (\ref{eq.NonConstDielectric}) is easily identified as a Coulomb interaction with a $q$-dependent permittivity $\epsilon(q) = 1 + 2\pi \alpha_{2D}q$, which is consistent with the space-dependent permittivity suggested in the simplistic picture of Fig. \ref{fig:Exciton1}.

It is convenient to take the inverse Fourier transform of $\Phi_{2D}(\vec q)$, since one is normally interested in using it as the electron-hole interaction potential in an effective mass model (Schr\"odinger) equation in real-space. Finally, this results in 
\begin{equation}\label{eq.potentialfinal}
\Phi(\rho) = \frac{e}{4\alpha_{2D}} \left[ H_0 \left( \frac{\rho}{r_0}\right) - Y_0 \left(\frac{\rho}{r_0}\right)\right]
\end{equation}
where $H_0$ and $Y_0$ are the zeroth order Struve and Neumann functions, respectively, and the screening length is defined as $r_0 = 2\pi\alpha_{2D}$.

So far, the interaction potential has been developed for a suspended layer, i.e. surrounded by vacuum both above and below the sample. Most of the 2D materials investigated in the literature are however deposited on a substrate. In fact, the original derivation of Eq. (\ref{eq.potentialfinal}) by Keldysh \cite{29} was done for a slab with finite thickness $d$ and dielectric constant $\epsilon$, on top of a semi-infinite medium of dielectric constant $\epsilon_2$. Nevertheless, this derivation leads to the same potential as in Eq. (\ref{eq.potentialfinal}), but slightly modified, with $r_0 \rightarrow d\epsilon/(1+\epsilon_2)$. \cite{30}

\subsubsection{Wannier-Mott exciton Hamiltonian}

Once the interaction potential is known, the exciton binding energy can be calculated within the Wannier-Mott model, where the exciton wave function is assumed to cover a large region of the material, involving several atoms. In this case, one can treat electrons and holes as effectively free quasi-particles in conduction and valence bands, respectively, interacting via the screened Coulomb potential in Eq. (\ref{eq.potentialfinal}). In this context, the exciton Hamiltonian reads \cite{31}
\begin{eqnarray}\label{eq.ExcHamIni}
H_{exc} = -\frac{\hbar^2}{2}\sum_{ i = e,h } \left({\frac{1}{m^i_x}} {\frac{\partial^2 }{ \partial x^2_i } + \frac{1}{m^i_y}} {\frac{ \partial^2 }{\partial y^2_i }}\right) \quad \quad\quad\quad   \nonumber \\
 \quad \quad\quad\quad  + V_{eh}( |x_e - x_h|, |y_e-y_h|),
\end{eqnarray}
where electron and hole effective masses $m^{e}_{x/y}$ and $m^{h}_{x/y}$ are identified as those for conduction ($m^{+}_{x/y}$) and valence ($m^{-}_{x/y}$) band states in Eq. (\ref{eq.mass}), respectively, and $V_{eh} = -e\Phi$.  This four dimensional problem can be simplified by using the relative $\vec \rho = \vec{\rho}_e - \vec{\rho}_h$ and center-of-mass $\vec R = \left(m^e_x x_e {\hat x}_e + m^h_x x_h {\hat x}_h \right)/M_x + \left(m^e_y y_e {\hat y}_e + m^h_y y_h {\hat y}_h \right)/M_y$ coordinates system, where $M_{x(y)} = m^e_{x(y)} + m^h_{x(y)}$. After some algebra on the derivatives in Eq. (\ref{eq.ExcHamIni}), the exciton Hamiltonian becomes
\begin{equation}\label{eq.excitonFinal}
H_{exc} = -\frac{\hbar^2}{2\mu_x} \frac{\partial^2 }{\partial x^2 } -\frac{\hbar^2}{2\mu_y} \frac{\partial^2 }{\partial y^2 } + V_{eh}(x, y),
\end{equation}
where the center of mass terms were dropped, since they commute with $H_{exc}$ and thus are constants of motion.

\subsubsection{Exciton eigenstates}

\begin{figure}[!h]
\includegraphics[width = \linewidth]{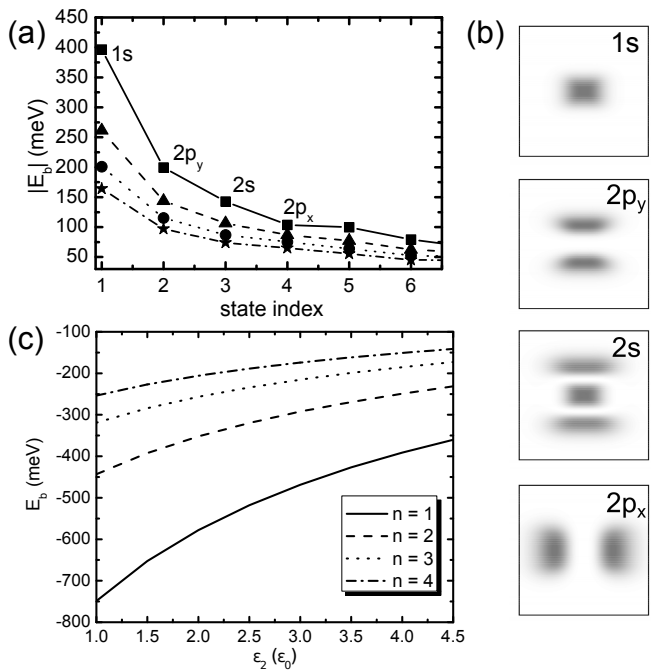}
\caption{(a) Exciton spectrum for $n$ = 1, 2, 3 and 4 layers BP, assuming a SiO$_2$ substrate ($\epsilon_2 = 3.9\epsilon_0$). (b) Wave functions for the 4 low-lying energy states in the monolayer case. (c) Ground state exciton energy of $n$ = 1, 2, ... 4 layers BP as a function of the dielectric constant of the substrate. Lines in (a) and (c) are guides for the eyes.}
\label{fig:Exciton2}       
\end{figure}

By solving Eq. (\ref{eq.excitonFinal}), one obtains the exciton spectrum in black phosphorus, which is shown in Fig. \ref{fig:Exciton2}(a) for $n = 1$ (squares), 2 (triangles), 3 (circles), and 4 (stars) layers BP. The exciton eigenfunctions of the 4 lowest energy states of the monolayer case are shown in Fig. \ref{fig:Exciton2}(b), where the eigenfunctions symmetry is indicated. Although, in general, the electron-hole pair resembles a hydrogen atom, three main differences are observed as compared to the latter:  (i) the Rydberg series of $s$ states do not follow the $E_b = -R_y/(n-1/2)^2$ rule, where $R_y$ is the Rydberg energy, (ii) the $s$ and $p$ states are not degenerate, and (iii) $p_x$ and $p_y$ states are also not degenerate. (i) and (ii) are ubiquitous of all 2D materials, as a consequence of the non-Coulombic electron-hole interaction. The lift of $s$/$p$ degeneracy is related to the break of the SO(3) symmetry of the Coulomb potential, whereas the non-hydrogenic Rydberg series can be explained by a energy-dependent dielectric constant, in accordance to Eq. (\ref{eq.NonConstDielectric}). As for (iii), the $p_x$/$p_y$ degeneracy is lift as a direct consequence of the effective mass anisotropy of BP.

	Results in Fig. \ref{fig:Exciton2}(a) are calculated for a system deposited on a silica substrate, with dielectric constant $\epsilon_2 = 3.9\epsilon_0$. It is however important to investigate the dependence of the exciton ground state binding energy on the dielectric constant of the substrate, in order to predict corrections on the observed optical gap due to different substrates or even in the case of a suspended sample (i.e. $\epsilon_2 = \epsilon_0$). This is shown in Fig. \ref{fig:Exciton2}(c) for $n$ = 1 to 4 layers BP, where we observe that as the dielectric constant of the substrate increases, the binding energies decrease due to the stronger screening, yielding a redshift of excitonic peaks. Results for the case of a suspended sample agree well with those obtained by more sophisticated calculations involving GW approximation and Bethe-Salpeter equations. \cite{16}

\begin{figure}[!h]
\includegraphics[width = \linewidth]{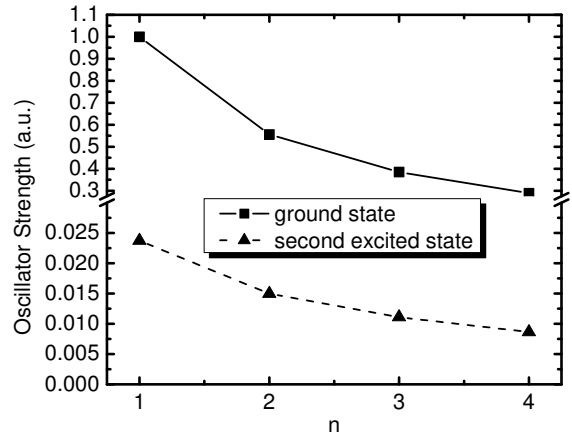}
\caption{Oscillator strength of the ground and second excited exciton states of few-layers BP, as a function of the number of layers. Values are relative to the ground state oscillator strength for monolayer BP.}
\label{fig:Exciton3}       
\end{figure}	
	
	From Fermi's golden rule, one obtains the transitions between exciton levels that are allowed to occur. Absorbed and emitted light in processes mediated by excitons is shown to be polarized along $x$ (armchair) direction, which is consistent with experimental results, as stated in previous Sections. Moreover, in usual absorption experiments, which involve one-photon processes, only $s$ states are optically active. The oscillator strength is then given by the square modulus of the exciton envelope function at $(x = 0, y = 0)$, which is non-zero for $s$ states. \cite{31} The oscillator strength of the ground and second excited exciton states of few-layer BP, which are demonstrated to be $s$ states in Figure \ref{fig:Exciton2}(b), are shown in Fig. \ref{fig:Exciton3} as a function of the number of layers. Notice the oscillator strength for the second excited state is almost two orders of magnitude lower than that of the ground state, which makes this state much harder to be experimentally probed. Besides, as the number of layers increase, the electron-hole interaction becomes weaker, leading to wider envelope functions and, consequently, to lower oscillator strengths.

\section{Thermal properties}

The thermal properties of black phosphorus (BP) are discussed in this section. As will be shown, the particular puckered atomic structure of the P atoms in BP leads to unusual thermal properties that are unique to this material and that can be beneficial for certain applications related to thermal management and thermoelectric conversion.

\subsection{Phonons}
\subsubsection{Phonon dispersion}

\begin{figure}[!h]
\includegraphics[width = \linewidth]{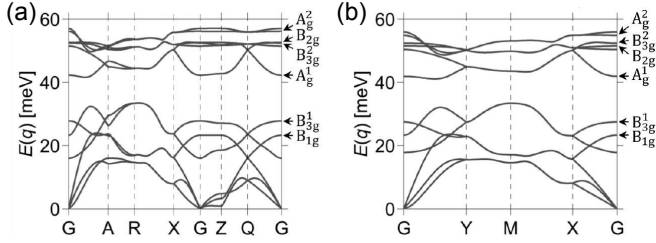}
\caption{Phonon dispersion along high symmetry $q$-points for (a) bulk and (b) monolayer BP. The Raman active optical phonon modes are labelled.}
\label{fig:Phonon1}
\end{figure}

Phonons, the discrete quanta of atomic vibrational excitations, are dominantly responsible for the thermal properties of BP. Using first principles modelling to extract the force constants characterizing the interatomic potential, the dynamical matrix is diagonalized to obtain the phonon energies \cite{32}, which are plotted for bulk and monolayer BP in Figures \ref{fig:Phonon1}(a) and \ref{fig:Phonon1}(b), respectively along the high symmetry points in the Brillouin zone previously shown in Figure \ref{fig.Crystal}. Each atom in the primitive cell has three degrees of freedom, leading to 12 phonon modes: 3 acoustic and 9 optical. In bulk BP, the three acoustic modes (one longitudinal and two transverse) have a linear dispersion in the vicinity of the Brillouin zone center, while in monolayer BP, the cross-plane polarized acoustic mode (ZA mode) has a quadratic dispersion, as is typical in 2D materials \cite{33}. Several theoretical reports have investigated the phonon properties of black phosphorus \cite{33, 34, 35, 36, 37, 38}.

From the slopes of the longitudinal acoustic phonon dispersion, one can extract the sound velocities along zigzag and armchair direction, giving 8429 m/s (zigzag) and 4313 m/s (armchair) for bulk BP, and 7955 m/s (zigzag) and 4037 m/s (armchair) for monolayer BP. The anisotropic sound velocity is directly related to the particular crystal structure of BP, and is similar for both bulk and monolayer. The acoustic modes travel at higher speed along the zigzag direction since all bonds are in the same plane as the phonon motion, and thus ``stiffer'' compared to the armchair direction where half the atomic bonds are nearly perpendicular (cross-plane) to the direction of phonon transport. The angle dependence in sound velocities suggests that the in-plane thermal transport properties will also be anisotropic. Figure \ref{fig:Phonon1} also labels which phonon modes at the G point are Raman active. Raman scattering experiments can be used, for example, to determine the crystal orientation of BP samples, probe local temperatures and measure induced strains \cite{37, 39}.

\subsubsection{Phonon transport}

To calculate the phonon transport properties of BP, the Landauer approach \cite{40} is employed. Within the Landauer formalism for solving the Boltzmann transport equation, there are two main quantities of interest: the number of conducting phonon modes (or channels), $M_{ph}$, and the phonon mean-free-path for backscattering, $\lambda_{ph}$. $M_{ph}$ is proportional to the product of the average velocity in the direction of transport $\langle |v_x| \rangle$ (transport is assumed to be along the $x$-direction) and the density of states $D$:
\begin{equation}
M_{ph} = \frac{h}{2} \langle |v_x| \rangle D
\end{equation}
where $h$ is Planck's constant. It depends only on the phonon dispersion of BP, and can be efficiently computed using the so-called “band counting” algorithm \cite{41} as implemented in the LanTraP simulation tool \cite{42}. $\lambda_{ph}$ is defined as the average distance travelled by a phonon along the transport direction before scattering from a forward-moving state to a reverse-moving state, which depends on the phonon dispersion and the scattering physics:
\begin{equation}
\lambda_{ph} = \frac{2 \langle v_x^2 \tau \rangle}{\langle |v_x| \rangle} 
\end{equation}
where $\tau$ is the phonon relaxation time.

Note that $M_{ph}$ and $\lambda_{ph}$ are only functions of energy, and that all averages are defined as $\langle A \rangle = \sum_k A(k) \delta (E-E_k)/\sum_k \delta(E-E_k)$, where $k$ is a reciprocal wave-vector in the Brillouin zone. The product of $M_{ph}$ and $\lambda_{ph}$ is equivalent to commonly-encountered solutions of the Boltzmann transport equation, sometimes referred to as the transport distribution \cite{40}, which is proportional to $\sum_k v_x^2(k) \tau(k) \delta(E-E_k)$

\begin{figure}[!h]
\includegraphics[width =\linewidth]{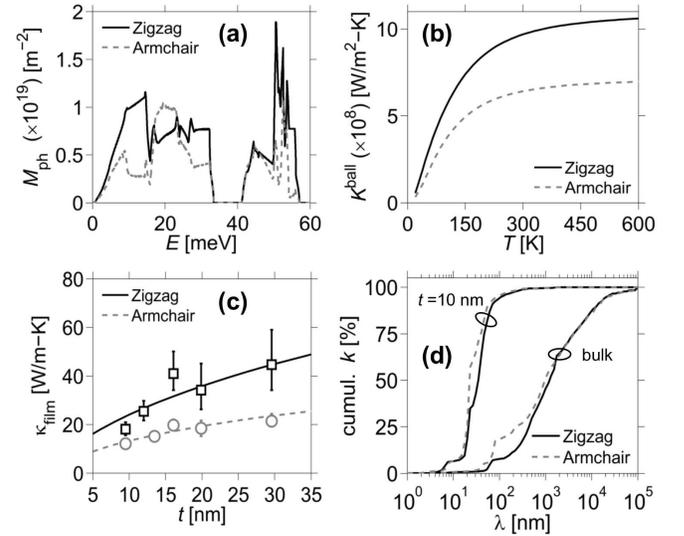}
\caption{(a) Number of phonon modes per cross-section, $M_{ph}$, of bulk BP versus energy, $E$. (b) Ballistic thermal conductance, $K^{ball}$, versus temperature, $T$. (c) In-plane thermal conductivity of thin BP films, $\kappa_{film}$, versus film thickness, $t$. Markers are experimental data and lines are theoretical calculations. Adapted from Ref. \onlinecite{39}. (d) Cumulative thermal conductivity versus phonon mean-free-path for backscattering, $\lambda_{ph}$, for bulk BP and a 10 nm-thick BP film}
\label{fig:Phonon2}
\end{figure}

Figure \ref{fig:Phonon2}(a) shows the number of phonon modes per cross-section, $M_{ph}$, versus phonon energy, $E$, for bulk BP with transport along zigzag and armchair directions. $M_{ph}$ is larger along the zigzag direction compared to armchair direction, particularly for the low-energy acoustic modes (less than 18 meV), which typically carry most of the heat. The angle dependence in the phonon channels is related to the anisotropy in the group velocities, which is larger along zigzag compared to armchair. Each phonon mode carries one quantum of thermal conductance, $K_0 = (\pi k_B)^2 T / 3h$, where $k_B$ is Boltzmann’s constant and $h$ is Planck’s constant. From the number of modes, $M_{ph}$, the ballistic thermal conductance, $K^{ball}$, can be calculated using \cite{40}: 
\begin{equation}
K^{ball} = K_0 \int_0^{\infty} M_{ph}(E) W_{ph}(E) dE
\end{equation}
where $W_{ph}(E) = 3(E/\pi k_B t)^2 [-\partial n_{BE}\partial E]$  is the so-called window function that depends on the Bose-Einstein occupation function, $n_{BE}$, and is normalized so that $\int_0^{\infty} W_{ph}(E) dE = 1$. Figure \ref{fig:Phonon2}(b) shows $K^{ball}$ as a function of temperature $T$, which corresponds to the thermal conductance of phonons in the ballistic transport limit (a quantity that is independent of the length of the sample). $K^{ball}$ is roughly 1.5$\times$ larger along zigzag versus armchair direction and increases monotonically as more phonon modes become thermally active at higher temperature. 

Thermal conductivity $\kappa$ is obtained in a similar way as $K^{ball}$, but including the average phonon mean-free-path for backscattering in the energy integral\cite{40}:
\begin{equation}
K^{ball} = K_0 \int_0^{\infty} M_{ph}\lambda_{ph}(E) W_{ph}(E) dE.
\end{equation}
Obtaining the phonon mean-free-path distribution requires knowing the dominant scattering physics. Typically, in bulk materials, the dominant scattering mechanisms for phonons are phonon-phonon (Umklapp) scattering, boundary scattering and defect scattering \cite{40}, while for thin films, surface scattering can become important \cite{43}. First principles calculations of thermal conductivity in monolayer BP, obtained from solutions of the Boltzmann transport equations including intrinsic three-phonon Umklapp scattering rates, predict notable anisotropy, with $\kappa$ in the range 30-110 W/m-K and 14-36 W/m-K for transport along the zigzag and armchair directions, respectively. To date, there are no experimental results on the thermal transport properties of monolayer BP. However, recent studies have probed the thermal transport properties of BP thin films \cite{39,44,45} with sub-micron thicknesses down to 10 nm. The in-plane measured thermal conductivities show anisotropic ratios in the range of 1.5 - 2.0, with zigzag being the preferential transport direction. In addition, $\kappa$ is found to increase with increasing film thickness, a signature of surface scattering of phonons. To model the in-plane thermal transport properties of the BP thin films, Umklapp phonon-phonon scattering and surface roughness scattering are included, both using phenomenological models and parameters described in Ref. \onlinecite{39}. 

Figure \ref{fig:Phonon2}(c) presents the in-plane film thermal conductivity $\kappa_{film}$ versus BP film thickness, $t$, as a function of transport direction (markers: experiment; lines: theory). Theoretical modeling results are found to compare well to the measurements, capturing both the anisotropy and the thickness dependence of $\kappa_{film}$. According to the model, the observed anisotropy in $\kappa_{film}$ originates from $M_{ph}$ (or $K^{ball}$), as opposed to strong angle-dependent phonon scattering. To understand how the different phonons contribute to the thermal conductivity, Figure \ref{fig:Phonon2}(d) shows the cumulative phonon thermal conductivity as a function of the phonon mean-free-path for backscattering, $\lambda_{ph}$. Comparing the case of bulk BP to that of a 10 nm-thick BP film, we find that surface scattering significantly reduces $\lambda_{ph}$ and, in particular, the phonons with the longest $\lambda_{ph}$. As an example, in a 10 nm-thick BP film, roughly 90\% of the heat is carried by phonons with $\lambda_{ph}$ less than 100 nm, compared to only 10-20\% in bulk BP. Such plots are also useful for identifying the phonons that contribute most to $\kappa$ (points of steepest slope), which can be selectively scattered in cases where low $\kappa$ is desired, for example in thermoelectric applications. \cite{He, Biwas}

\subsection{Thermoelectricity}

By coupling the phonon transport properties, described in the previous section, with the electronic transport properties, we can evaluate the thermoelectric (TE) performance of BP. The unusual in-plane transport of BP is found to be beneficial for TE conversion. The efficiency of TE conversion, that is using a heat source to generate electrical power or conversely by applying an electrical current to achieve cooling, is characterized by the TE figure-of-merit, \textit{ZT}, which can span from zero to infinity (the Carnot efficiency is reached as $ZT \rightarrow \infty$). \textit{ZT} is written as $S^2\sigma T/\kappa$, where $S$ is the Seebeck coefficient, $\sigma$ the electrical conductivity, $T$ the temperature and $\kappa = \kappa_{el} + \kappa_{ph}$ is the thermal conductivity with contributions from both electrons and phonons. The goal is to increase $S^2\sigma$, also known as the power factor, \textit{PF} (which has historically proven difficult), and to decrease $\kappa$ (which has led to record \textit{ZT} values in the past decade). Several, mostly theoretical, studies have investigated the thermoelectric properties of BP \cite{26,48,49,50,51,52}, which shows an interesting anisotropic behavior, as we will discuss below.

\begin{figure}[!h]
\includegraphics[width = \linewidth]{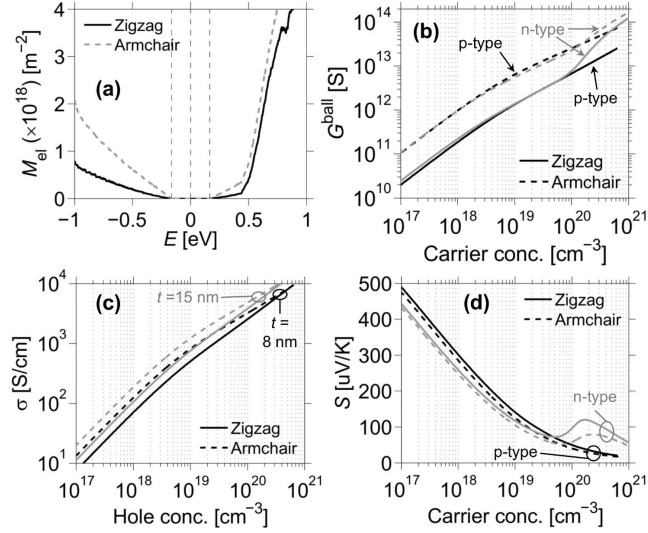}
\caption{(a) Number of electron modes per cross-section, $M_{el}$, of bulk BP versus energy, $E$. (b) Ballistic conductance, $G^{ball}$, versus carrier concentration. (c) Electrical conductivity $\sigma$ versus hole carrier concentration for 8 nm- and 15 nm-thick BP films. (d) Seebeck coefficient, $S$, versus carrier concentration. $T$ = 300 K.}
\label{fig:Phonon3}
\end{figure}

Using the Landauer approach, TE parameters can be calculated, allowing one to assess the TE performance of BP \cite{46}. Similarly to the phonon transport properties, the electron transport properties depend on two main quantities: the number of conducting electron modes (or channels), $M_{el}$, and the electron mean-free-path for backscattering, $\lambda_{el}$. $M_{el}$ depends only on the electron band structure, while $\lambda_{el}$ depends on the particular scattering physics, as well as on the electron dispersion. Note that the expressions for the number of electron modes, $M_{el}$, and the electron mean-free-path for backscattering, $\lambda_{el}$, are identical to those for phonons (i.e. $M_{ph}$ and $\lambda_{ph}$). Figure \ref{fig:Phonon3}(a) presents the number of electron modes per cross-section, $M_{el}$, versus energy, $E$, for bulk BP with transport along zigzag and armchair directions. $M_{el}$ is largest with transport along the armchair direction, due to the larger effective mass along the zigzag direction. BP possesses unusual in-plane anisotropy, where the preferential transport directions are orthogonal for electrons (armchair) and phonons (zigzag). This orthogonal anisotropy is beneficial for TE operation since \textit{PF} is maximized and $\kappa_{ph}$ is minimized ($\kappa_{ph}$ is larger than $\kappa_{el}$ in BP) with transport along armchair. Each electron mode carries one quantum of conductance, $2e^2/h$ , where $e$ is the electron charge. From the number of modes $M_{el}$, the ballistic conductance, $G^{ball}$, can be straightforwardly calculated \cite{47}. Figure \ref{fig:Phonon3}(b) shows $G^{ball}$ versus carrier concentration, which corresponds to the conductance of electrons in the ballistic transport limit (a quantity that is independent of the length of the sample). $G^{ball}$ is significantly larger, roughly 5$\times$, with transport oriented along the armchair direction.
	
Electrical conductivity $\sigma$ is obtained by multiplying $G^{ball}$ by the average electron mean-free-path for backscattering \cite{46}. Typically in bulk TE materials the dominant scattering mechanisms for electrons are electron-phonon scattering and impurity scattering, as well as surface roughness scattering for thin films. First principles-based calculations of electron-acoustic phonon scattering within the deformation potential approach have predicted electron/hole mobilities as high as thousands to tens of thousands of cm$^2$V$^{-1}$s$^{-1}$ in few-layer BP, with anisotropy ratios greater than 30 in monolayer BP \cite{15}. Regarding the TE properties of bulk, monolayer, nanoribbon, alloyed and strained BP, theoretical calculations have predicted optimal \textit{ZT} values over a wide range from 0.72 to 6.4 \cite{26, 48, 49, 50, 51}. For comparison, state-of-the-art TE materials have \textit{ZT} greater than unity with record values in the range 2-3. Experimentally few works have probed the TE characteristics of BP \cite{52}. Here, the TE performance of few-layer BP is assessed by calibrating our theoretical model to existing experimental data. Using the measured angle-dependent electronic mobility of few-layer BP \cite{1,5,7} we find $\lambda_{el}$ is equal to 38 nm (zigzag) and 13 nm (armchair) for a 8 nm-thick BP film, and 56 nm (zigzag) and 19 nm (armchair) for a 15 nm-thick BP film \cite{1}. Note that BP is experimentally p-type, so that the present model is thus calibrated for hole conduction. 

Figure \ref{fig:Phonon3}(c) shows the electrical conductivity $\sigma$ versus hole concentration for BP films of thickness 8 nm and 15 nm. The anisotropy ratio in $\sigma$ is roughly 1.5 - 2.0, which is less than that of $G^{ball}$. The thicker film has higher $\sigma$, due to reduced surface scattering. Figure \ref{fig:Phonon3}(d) presents the Seebeck coefficient, $S$, as a function of carrier concentration (with an energy-independent $\lambda_{el}$, $S$ does not depend on $\lambda_{el}$ or film thickness). $S$ is found to be relatively insensitive to BP orientation. Interestingly, at high carrier concentration there is a peak in $S$ for n-type BP, and an associated increase in $G^{ball}$, due to a secondary band. This suggests that BP could potentially be a better n-type TE material, as opposed to p-type.

\begin{figure}[!h]
\includegraphics[width = \linewidth]{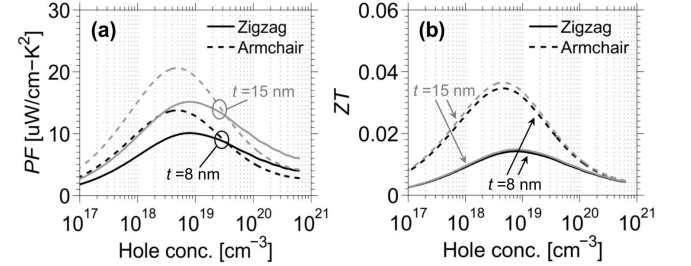}
\caption{(a) Thermoelectric power factor $PF = S^2\sigma$ versus hole concentration for 8 nm- and 15 nm-thick BP films. (b)Thermoelectric figure-of-merit, \textit{ZT}, versus hole concentration for 8 nm- and 15 nm-thick BP films. T = 300 K.}
\label{fig:Phonon4}
\end{figure}

Having calculated all the TE parameters (both electron and phonon), which are calibrated to available experimental data, we can assess the TE potential of few-layer BP. Figure \ref{fig:Phonon4}(a) shows the power factor, $PF = S^2\sigma$, versus hole concentration for BP films of thickness 8 nm and 15 nm. \textit{PF} reaches 20 $\mu$W/cm-K$^2$, which is a respectable value when compared to good TE materials such as Bi$_2$Te$_3$ ($\approx$30 $\mu$W/cm-K$^2$). The anisotropy is similar to that in $\sigma$. The thicker film achieves a larger \textit{PF} due to reduced surface roughness scattering. The TE figure-of-merit, \textit{ZT}, is presented in Figure \ref{fig:Phonon4}(b) as a function of hole concentration and BP film thickness. The anisotropy ratio reaches 3 for \textit{ZT}, with the highest \textit{ZT} along the armchair direction, which results from the orthogonal preferential transport directions for electrons and phonons. \textit{ZT} is nearly thickness insensitive, since surface roughness scattering reduces both $\sigma$ (numerator of \textit{ZT}) and $\kappa_{ph}$ (denominator of \textit{ZT}). The maximum \textit{ZT} at room temperature approaches 0.04 with a hole concentration of 5$\times$10$^{18}$ cm$^{-3}$. While good TE materials should have \textit{ZT} above unity, these results are calibrated to some of the first experimental reports and should improve as the approaches and material quality improves. By comparison, other 2D materials (monolayer and few-layer) have experimentally shown large \textit{PF} values, such as 85 $\mu$W/cm-K$^2$ in few-layer MoS$_2$ \cite{53}, but large thermal conductivities in the range of 50 W/m-K \cite{54} keep \textit{ZT} low.

\section{Mechanical Properties-elasticity}

\begin{figure}[!h]
\includegraphics[width = \linewidth]{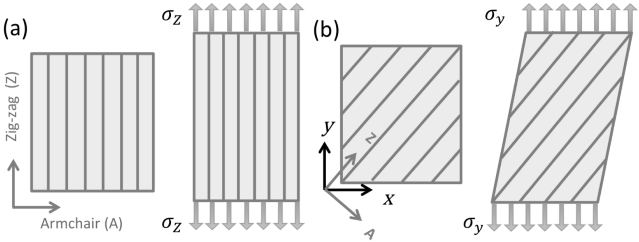}
\caption{Schematics of the deformation of an orthotropic material upon the application of stress along (a) the zig-zag (Z) preferential and (b) an arbitrary ($y$) direction. The $xy$ Cartesian system associated to this direction makes an angle $\theta$ with the preferential AZ Cartesian system.}
\label{fig:Elasticity1}
\end{figure}

The structural anisotropy of BP is also manifested in a highly-directional dependent elastic behavior. The in-plane linear elastic behavior can be modeled with an orthotropic continuum plate model, with the principal axes oriented along the armchair (A) and zig-zag (Z) directions. Figure \ref{fig:Elasticity1}(a) shows the deformation along one preferential direction, such as the Z direction. Upon application of unit stress $\sigma_Z$, \footnote{Not to confuse with the conductivity $\sigma_i$ in the $i$-direction ($i = x, y, z$), discussed in previous sections. From here onwards, the variable $\sigma_i$ denotes \textit{stress} in the $i$-direction.} the plate elongates along Z while shrinking in the A direction. We characterize this deformation by the strains $\varepsilon_Z = 1/E_Z$ and $\epsilon_A = -\nu_Z\varepsilon_Z$. Here $E_Z$ and $\nu_Z$ are the Young’s modulus and Poisson’s ratio associated to the Z direction, respectively. Figure \ref{fig:Elasticity1}(b) describes the deformation in the general case, in which the stress is applied along the arbitrary $y$ direction, which makes an angle $\theta$ with the Z direction. The orthotropic plate captures the shear-extension coupling, i.e. the applied $\sigma_y$ leads not only to extension strain along $y$ ($\varepsilon_y$) and compression strain along $x$ ($\varepsilon_x$), but also to shear deformation ($\varepsilon_{yx}$). The general stress-strain relation writes
\begin{equation}\label{eq.orthotropic}
\left(\begin{tabular}{c}
$\varepsilon_y$\\
$\varepsilon_x$\\
$\varepsilon_{yx}$
 \end{tabular} \right)  = \left(\begin{tabular}{ccc}
$\frac{1}{E_x}$ & $-\frac{\nu_y}{E_y}$ & $\frac{\eta_x}{G_{yx}}$ \\
$-\frac{\nu_x}{E_x}$ & $\frac{1}{E_y}$ & $\frac{\eta_y}{G_{yx}}$\\
$\frac{\eta_x}{G_{yx}}$ & $-\frac{\eta_y}{G_{yx}}$ & $\frac{1}{G_{yx}}$
 \end{tabular} \right)\left(\begin{tabular}{c}
$\sigma_y$\\
$\sigma_x$\\
$\sigma_{yx}$
 \end{tabular} \right) 
\end{equation}
where the Young's ($E_x$ and $E_y$), shear ($G_{yx}$) moduli, and Poisson's ratios ($\nu_x$ and $\nu_y$) associated to the $xy$ Cartesian system relate to the corresponding set of values along the principal axis via closed form expressions given elsewhere. \cite{55} The shear-strain coupling coefficients $\eta_x$ and $\eta_y$ are directly proportional to $\sin\theta$, so that they vanish when the $xy$ Cartesian system coincides with the principal one. 

The orthotropic model reduces to an isotropic plate model when $E_A = E_Z = E$, $\nu_A = \nu_Z = \nu$ and the shear modulus writes $G = E/2(1+\nu)$. The coupling coefficients $\eta_x$ and $\eta_y$ vanish in this case, and rotation of Cartesian coordinates system yields no change in the stress-strain relation. If we apply unit stress in $y$-direction (i.e. $\sigma_y = 1$ and $\sigma_x = \sigma_{yx} = 0$) then consequent strains would be $\varepsilon_x = 1/E$ (extension), $\varepsilon_y = -\nu\varepsilon_x$ (compression), and $\varepsilon_{yx} = 0$. The isotropic model is suitable for modeling e.g. the in-plane deformation of graphene \cite{56,57} and blue phosphorene. \cite{Unpublished}
	
	The set of elastic constants along the principal axes can be calculated from atomistic-level calculations. \textit{Ab initio} calculations obtained that for the monolayer BP, the Young's modulus of 166 GPa in the Z direction is 3.8 times larger than its counterpart in the armchair direction. For three-layer BP, its value in the Z direction, 160 GP, is 4.3 times larger than its counterpart in the armchair direction. \cite{58} The shear modulus for the monolayer $G_{ZA} = G_{AZ}$ was found to be 41 GPa.  We note that compared to other 2D materials, such as graphene and MoS$_2$, BP presents a smaller Young's modulus. For example, the reported Young's moduli for graphene and MoS$_2$ are 1.0 TPa \cite{59} and 0.25 TPa \cite{60}, respectively. This smaller Young's modulus in BP may be the result of the weaker P-P bond strength and the significant deformation of dihedral angles rather than bond length stretch. For the same reason, the shear modulus of BP is an order of magnitude smaller than the 472 GPa shear modulus of graphene. \cite{56} 
	
As in the case of graphene, there is an uncertainty on the effective value of the layer thickness. Thus, it is important to note that Young’s and shear modulus values mentioned above incorporate a 5.5 \AA\, thickness value for the monolayer and 16.65 \AA\, for the three-layer BP. \cite{60}  
	
\begin{figure}[!h]
\includegraphics[width = \linewidth]{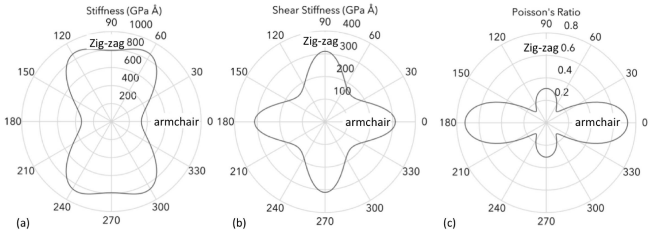}
\caption{(a) Surface Young's, (b) shear modulus, and (c) Poisson's ratio of BP as a function of direction. From Ref. \onlinecite{59}.}
\label{fig:Elasticity2}
\end{figure}

Atomistic calculations also obtained that BP exhibits an anisotropic Poisson's ratio. Its value along the Z direction (0.62 from Ref. \onlinecite{58}, 0.93 from Ref. \onlinecite{61}) is 2 - 4 times larger than that in the A direction (0.17 from Ref. \onlinecite{58} 0.4 from Ref. \onlinecite{61}). Figure \ref{fig:Elasticity2} presents the directional dependence of the elastic constants and Poisson's ratio as predicted with the orthotropic model and the elastic constants obtained from a tight-binding potential. \cite{62} Note that in order to avoid the plate thickness ambiguity mentioned above, the theoretical calculations summarized in Figure \ref{fig:Elasticity2}(a-b) are presenting the directional dependence of the surface Young's ($E_s$) and surface shear moduli ($G_s$) \cite{Unpublished}. From these plots, we observe that the relation $E_s = G_s /[2(1+\nu)]$ does not hold, meaning that an isotropic membrane model cannot be associated to BP. The information summarized in Fig. \ref{fig:Elasticity2} can be used with Eq. (\ref{eq.orthotropic}) of the orthotropic plate model to describe the in-plane deformation of BP along an arbitrary direction.

Black phosphorus presents anisotropic bending flexibility, as illustrated in Fig. \ref{fig:Exciton3}. The bending constant along the Z direction, of 1,300 GPa \AA\,$^3$, is about 2.5 larger that along the A direction. These values are larger than the bending modulus of graphene \cite{57}, of 230 GPa \AA\,$^3$. The anisotropy in bending impacts the formation of out-of-plane ripples under compressive strains. First-principles calculations demonstrated that BP develops ripples along the Z direction, \cite{63} where the out-of-plane deformation effectively releases the compressive strain energy, on one hand. Compression-induced deformation along the armchair direction is dominated by bond-angle distortion without any appreciable ripple formation, on the other hand. 

\begin{figure}[!h]
\includegraphics[width = 0.9\linewidth]{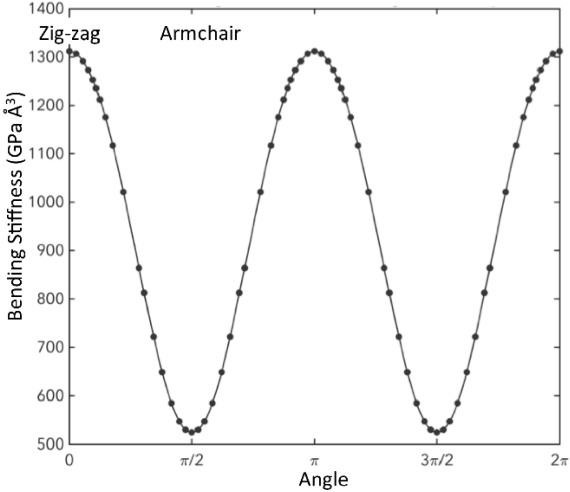}
\caption{Bending stiffness as a function of direction angle in black phosphorus.}
\label{fig:Elasticity3}
\end{figure}

Moving beyond linear elasticity, theoretical \textit{ab initio} calculations obtained that BP can sustain tensile strains up to 27\% and 30\% in the zigzag and armchair directions, respectively. \cite{58} A pronounced difference in the deformation modes under different strain direction has been identified: While tensile strain in the zigzag direction induces a larger P–P bond elongation, tensile strain applied in the armchair direction stretches BP without significantly extending the P–P bond length. Since first principles calculations are performed at 0 K, recent molecular dynamics simulations examined phosphorene's fracture behavior at finite temperatures. Temperature plays a significant role, with the fracture strength and strain diminishing by more than 65\% when increasing the temperature from absolute zero (0 K) to 450 K. Thermal sensitivity also depends on the zig-zag or armchair direction. \cite{64} At room temperature, the fracture strain of the defect-free BP in the zig-zag (armchair) direction is 13\% (6\%).

\section{Concluding remarks}

We have provided a general theoretical overview of the optical, electronic, thermoelectric, and mechanical properties of bulk and few-layer black phosphorus. All these properties show unusually large in-plane anisotropy that originate from BP's particular puckered atomic structure. These anisotropic properties makes BP an interesting material in the 2D family.

\acknowledgements AC acknowledges financial support from CNPq, through the Science Without Borders and PRONEX/FUNCAP programs, and from the Lemann Foundation. JM acknowledges support from NSERC (Discovery Grant RGPIN-2016-04881). TL acknowledges funding from the National Science Foundation through the University of Minnesota MRSEC under Award Number DMR-1420013.

\end{document}